\pdfoutput=1

\documentclass[9pt,twocolumn,twoside]{pnas-new}

\templatetype{pnasresearcharticle} 

\title{Universal Urban Spreading Pattern of COVID-19 and Its Underlying  Mechanism}

\author[a,1]{Yongtao Zhang}
\author[a,1]{Hongshen Zhang}
\author[a]{Mincheng Wu}
\author[a,2]{Shibo He}
\author[b]{Yi Fang}
\author[a]{Yanggang Cheng}
\author[c]{Zhiguo Shi}
\author[a]{Cunqi Shao}
\author[a]{Chao Li}
\author[d]{Songmin Ying}
\author[e]{Zhenyu Gong}
\author[b]{Yu Liu}
\author[b]{Xinjiang Ye}
\author[b]{Jinlai Chen}
\author[a]{Youxian Sun}
\author[a,2]{Jiming Chen}
\author[f,2]{H. Eugene Stanley}

\affil[a]{College of Control Science and Engineering, Zhejiang University, Hangzhou, China}
\affil[b]{Westlake Institute for Data Intelligence, Hangzhou, China}
\affil[c]{College of Information Science and Electronic Engineering, Zhejiang University, Hangzhou, China}
\affil[d]{School of Medicine, Zhejiang University, Hangzhou, China}
\affil[e]{Zhejiang Provincial Center for Disease Control and Prevention, Hangzhou, China}
\affil[f]{Center for Polymer Studies and Physics Department, Boston University, Boston, MA 02215, USA}

\leadauthor{Zhang} 

\significancestatement{By leveraging the trajectory data of 197,808 smartphone users (including 17,808 anonymous confirmed cases) in 9 cities in China, we find a universal spreading pattern in all cities: the spatial distribution of confirmed cases follows a power-law-like model and the spreading centroid is time-invariant. Moreover, we reveal that human mobility in a city drives the spatial-temporal spreading process: long average travelling distance results in a high growth rate of spreading radius and wide spatial diffusion of confirmed cases. Our results unveil the underlying mechanism behind the spatial-temporal urban evolution of COVID-19, and can be used to evaluate the performance of mobility restriction policies implemented by many governments and to estimate the evolving spreading situation of COVID-19.}

\authordeclaration{The authors declare no competing interests.}
\equalauthors{\textsuperscript{1}Y.Z. and H.Z. contributed equally to this work.}
\correspondingauthor{\textsuperscript{2}To whom correspondence should be addressed. E-mail: s18he@zju.edu.cn, cjm@zju.edu.cn, hes@bu.edu}


\begin{abstract}
Currently, the global situation of COVID-19 is aggravating, pressingly calling for  efficient control and prevention measures. 
Understanding spreading pattern of COVID-19 has been widely recognized as a vital step for implementing  non-pharmaceutical measures. Previous studies investigated such an issue in large-scale (e.g., inter-country or inter-state) scenarios  while urban spreading pattern still remains an open issue. Here, we fill this gap by leveraging the trajectory data of 197,808 smartphone users (including 17,808 anonymous confirmed cases) in 9 cities in China.
We find a universal spreading pattern in all cities: the spatial distribution of confirmed cases follows a power-law-like model and the spreading centroid is time-invariant.
Moreover, we reveal that human mobility in a city drives the spatial-temporal spreading process: long average travelling distance results in a high growth rate of spreading radius and wide spatial diffusion of confirmed cases. With such insight, we adopt Kendall model to simulate urban spreading of COVID-19 that can well fit the real  spreading process. Our results unveil the underlying mechanism behind the spatial-temporal urban evolution of COVID-19, and can be used to evaluate the performance of mobility restriction policies implemented by many governments and to estimate the evolving spreading situation of COVID-19.
\end{abstract}

\dates{This manuscript was compiled on \today}
\doi{\url{www.pnas.org/cgi/doi/10.1073/pnas.XXXXXXXXXX}}

\begin{document}

\maketitle
\thispagestyle{firststyle}
\ifthenelse{\boolean{shortarticle}}{\ifthenelse{\boolean{singlecolumn}}{\abscontentformatted}{\abscontent}}{}

\dropcap{A}s of Dec. 18, COVID-19 has struck over 222 countries, resulting in 72,851,747 cases and 1,643,339 deaths\cite{realtime2020}. The maximum of global daily new cases have exceeded 600,000, and situations in many countries are increasingly aggravating. It is pressing to find efficient ways to suppress the transmission of SARS-CoV-2\cite{bi2020epidemiology,linton2020incubation,li2020early,lauer2020incubation,kissler2020projecting,science:Ferrettieabb6936,NatureMedicine:2020}, which has been recognized as top priority since the very beginning of the initial outbreak in China and reiterated by the multidisciplinary online conference on Aug. 3, 2020, organized by the World Health Organization\cite{conference:who}.

Recently, resurgences of COVID-19 have been reported in many countries (e.g., United Kingdom, France, Spain). 
When a local resurgence takes place\cite{leonardo:naturehuman:reemergence:2020, Alberto:naturehuman:test:social:2020},  a fundamental issue for practical control and prevention is how does COVID-19 spread temporally and spatially within a city?  
Previous studies have investigated wide-range (e.g., inter-country or inter-state) spreading patterns of infectious diseases\cite{lancet:sun:epidemiological:2020,jia:populationflow:2020,pnas:transportationnetwork:prediction:2006,brockmann2013hidden}. They fall short for urban scenario, since the spreading of infectious diseases is highly dependent on spatial scale\cite{pnas:multiscalemobility:prediction:2009}. Moreover, the difficulty in obtaining fine-grained trajectory data of infectious cases has hindered the in-depth investigation of urban spreading pattern\cite{pei2018forecasting}. As such, little is known about the urban spreading pattern of COVID-19 so far, one of the most important spreading characteristics.

Here, we fill this gap by leveraging the trajectory data of 197,808 smartphone users (including 17,808 anonymous confirmed cases) in 9 cities in China. To have a comprehensive analysis, we select confirmed cases from Wuhan (where the initial outbreak of COVID-19 took place in China), Beijing and Urumqi (the cities with COVID-19 resurgences) and other cities (where cases were mainly imported). We try to address the following three questions: 1) Does the spreading of COVID-19 in different cities have a universal pattern? 2) What is the underlying mechanism for the spreading pattern of COVID-19? 3) How to utilize such a spreading pattern for control and prevention of COVID-19? 
We find a universal spreading pattern existing in all cities: the spatial distribution of confirmed cases follows a power-law-like model and the spreading centroid is time-invariant. 
Moreover, we reveal that human mobility in a city drives the spatial-temporal spreading process: long average travelling distance results in a high growth rate of spreading radius and wide spatial diffusion of confirmed cases. With such insight, we can accurately predict the shapes of spatial distribution of cases and the time when the peak of COVID-19 cases arrives. Our results  unveil the underlying mechanism behind the spatial-temporal urban evolution of COVID-19, and can be used to evaluate the performance of mobility restriction policies implemented by many governments and to estimate the evolving spreading situation of COVID-19.

\section*{Results}
\subsection*{Characterize the Activity Centroids of Confirmed Cases}
We adopt a collection of trajectory data contributed by anonymous smartphone users in China. The trajectory data record activity locations and corresponding timestamps when smartphone users are using location-based services. Clearly, these data reflect the real-time activity locations of smartphone users, at which they might get infected or infect others. To see how the spreading of COVID-19 is spatially correlated with the activity locations of confirmed cases, we first characterize their activity locations by a statistical metric: activity centroid (denoted by $\sigma$). The activity centroid of a smartphone user is defined as the average of activity locations reported by this smartphone user within a given period (e.g., 1 month). That is, assuming that there are $N_j$ activity locations $P^j=\left\{P_1^j,P_2^j,\cdots,P_{N_j}^{j}\right\}$ for a smartphone user $j$, the activity centroid $\sigma_j$ for user $j$, is defined as $\sigma_j =\mathbb{E}\left(P^j\right) = {\sum_k P_k^j}/{N_j}$.

\begin{figure*}
\centering
\includegraphics[width=17.5cm]{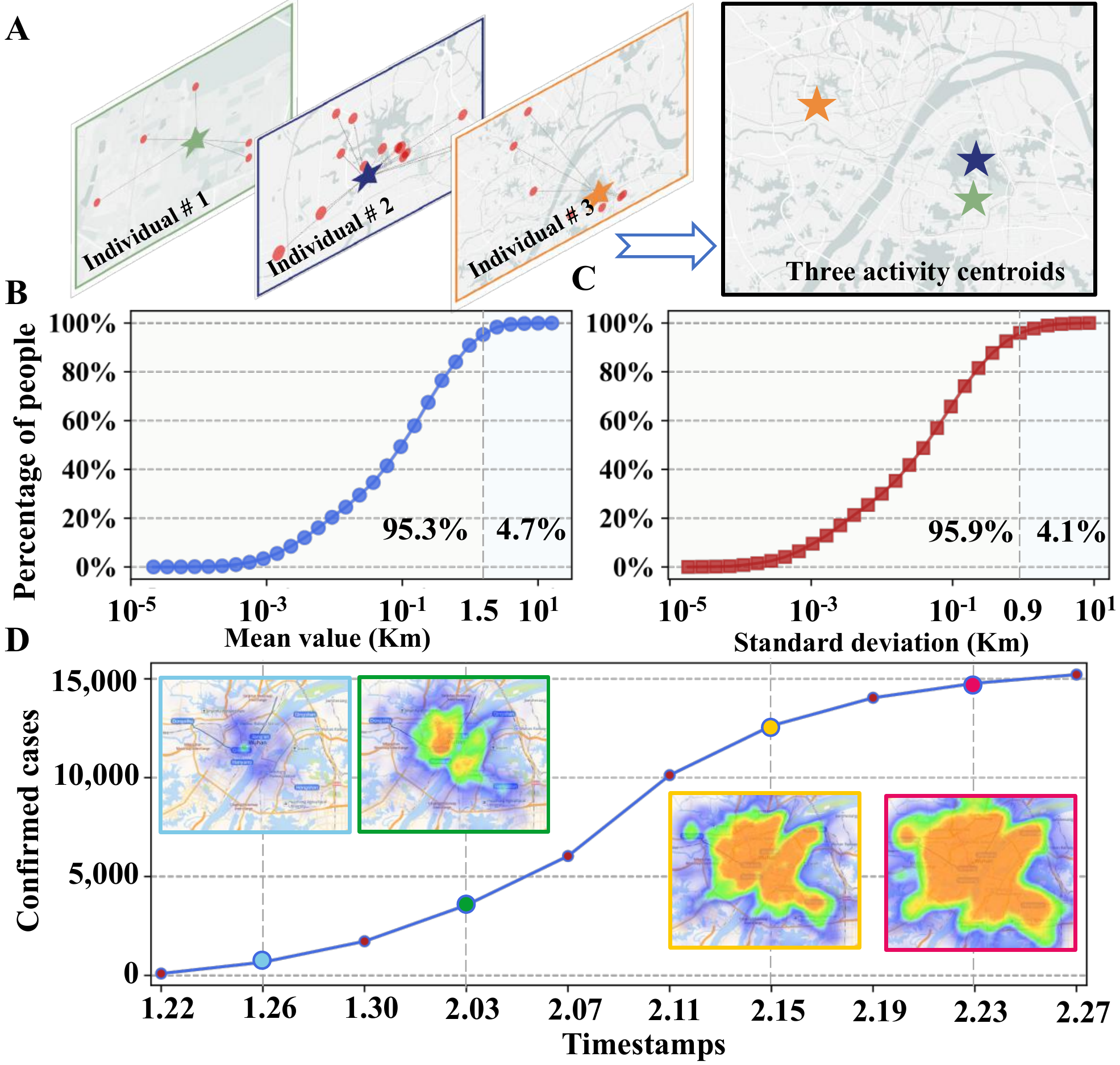}
\caption{Activity centroids and a visualization of COVID-19 spreading in Wuhan.
(\textit{A}) An illustration of calculating the activity centroids of smartphone users based on their activity locations. 
(\textit{B} to \textit{C}) The cumulative distributions of the mean and variance of the distance between activity centroids of each smartphone user in 1-6 months and the average centroid. 
(\textit{D}) Visualizing the spatial-temporal spreading of COVID-19 in Wuhan between January 22, 2020 and February 27, 2020.}
\label{fig:centroid}
\end{figure*}

We illustrate the calculation of activity centroid $\sigma$ in Fig. \ref{fig:centroid}A. The activity centroid should be stable over time for most smartphone users so that it can represent the intrinsic characteristic of activity locations. For this purpose, we randomly select 20,000 smartphone users in Wuhan and calculate their activity centroids in different periods (1 to 6 months), obtaining 6 activity centroids $\sigma_{j}(t),t\in[1,6]$ for each smartphone user $j$. The distances between 6 activity centroids and the average centroid (i.e., $\overline{\sigma_j}= \sum_{t=1}^6 \sigma_{j}(t)/{6}$) for each $j$ are then calculated. The mean and standard variance of these 6 distances for each user are computed and the cumulative distributions of all smartphone users are displayed in Figs. \ref{fig:centroid}B and \ref{fig:centroid}C. Clearly, the mean values of 95.3\% smartphone users are less than 1.5 kilometer ($Km$) and the standard variance is relatively small. This implies that the activity locations of smartphone users exhibit strong stability, independent of the chosen period (see sensitivity analysis of the periods in Fig. S2). Such an intrinsic behavior can be well captured by the activity centroid.

The most frequently visiting location (MVL) is also a statistical metric of great interest. It indicates the activity location that a smartphone user visits most frequently. We calculate the percentage of top $k$ ($k=1,2,\cdots,5$) activity locations of each smartphone user and show the average of all 20,000 smartphone users in Figs. S1A-S1C. The MVL (i.e., top 1 activity location) only accounts for about 45\% of all activity locations, that is, more than one half of activity location information is not utilized by MVL to characterize the activity behavior. Further, the performance of the metric of top $k$  activity locations  approaches that of activity centroid when $k$ increases (see detailed information in Figs. S1D-S1I). Therefore, we choose activity centroid as the statistical metric instead of MVL in this article.

By collaborating with Westlake Institute for Data Intelligence and local institutions for disease control and prevention, we obtain a dataset of confirmed cases who are also smartphone users for location-based services, including their activity centroids and dates of confirmation (see details on the data description in supplementary materials). We visualize the spreading process of COVID-19 in Wuhan through heat map in Fig. \ref{fig:centroid}D.
The introduction of activity centroid enables us to quantify the spatial and temporal spreading pattern of COVID-19 in the following sections.

\subsection*{The Temporal Spreading Pattern}
\begin{figure*}
\centering
\includegraphics[width=17.5cm]{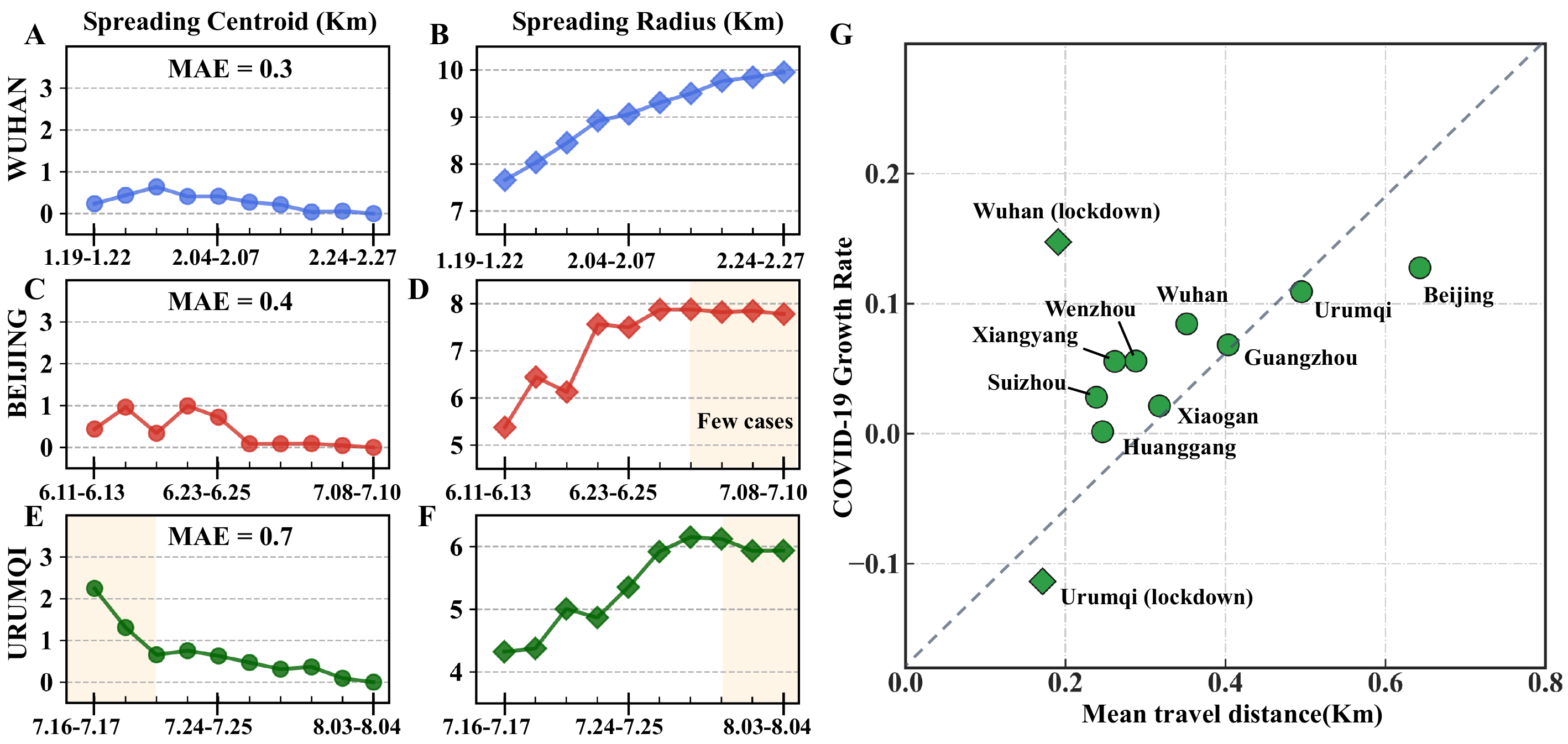}
\caption{ The temporal spreading pattern of COVID-19. 
(\textit{A} to \textit{F}) The cumulative spreading centroid and spreading radius in Wuhan, Beijing and Urumqi, respectively. 
(\textit{G}) Relation between mean travel distance of people in each city and corresponding COVID-19 growth rate.}
\label{fig:temporal:pattern}
\end{figure*}

To characterize the temporal spreading pattern, we divide the spreading duration in each city into $L$ equal periods, and allocate confirmed cases in set $U$ into subset $U_i$ if their confirmation dates are within $i^{th}$ period, $i\le L$. An illustration of division is provided in the x-axis index of Figs. \ref{fig:temporal:pattern}A, C and E,  where $L=10$. Given a set $U$ of all confirmed cases, we are able to calculate the overall spreading centroid (denoted by $\rho$) of $U$ as the average of their activity centroids, i.e., $\rho = \mathop{\mathbb{E}} (\sigma)$, where $\sigma=\{\sigma_j\}_{j\in U}$. The cumulative spreading centroid is then defined as corresponding cumulative value until $i^{th}$ period, that is, the averages of activity centroids of confirmed cases in set  $\bigcup_{k=1}^i U_k$. 

We display the difference between the overall spreading centroid and cumulative spreading centroid until $i^{th}$ period in Figs. \ref{fig:temporal:pattern}A, C and E for Wuhan, Beijing and Urumqi, respectively. Interestingly, as the situation of COVID-19 evolves, cumulative spreading centroids at different periods in Wuhan, Beijing and Urumqi are close to the overall spreading centroid: the Mean Absolute Errors (MAE) between cumulative and overall spreading centroids in Wuhan, Beijing and Urumqi are 0.3 $Km$, 0.4 $Km$ and 0.7 $Km$, respectively. It is clear to see that the temporal spreading centroid of COVID-19 has a feature of time invariance. That is, the spreading centroid is stable and nearly does not migrate during COVID-19 spreading. 

We now study the spreading radius $\gamma$ of set $U$, defined as $\gamma =  \sum_{j\in U}{d}(\sigma_j,\rho)/|U| $, where $d(\cdot)$ is the Euclidean distance between the activity centroid of a confirmed case and the spreading centroid  $\rho$ of a set $U$ of confirmed cases, and $|U|$ denotes the number of confirmed cases in $U$. Obviously, the spreading radius quantifies the mean distance between confirmed cases' activity centroids and the spreading centroid. Similarly, we introduce cumulative spreading radius for set $\bigcup_{k=1}^i U_i$. The cumulative spreading radius in different periods increase slowly over time in these three cities except for periods with few cases (less than 10 cases).

We also conduct analysis for the other 6 cities with relative large number of confirmed cases in China: Xiaogan, Suizhou, Xiangyang, Huanggang, Guangzhou and Wenzhou (Fig. S4), where cases are mainly imported. Similar conclusions for spreading centroid and spreading radius can be made in these cities. We proceed to perform sensitivity analysis by varying the number of spreading period $L$ (Fig. S3) and find that $L$ does not have much impact on the observed temporal pattern. Therefore, the temporal spreading pattern of COVID-19 in China features time invariance of spreading centroid and slow growth of spreading radius. 

As can be seen in Figs. \ref{fig:temporal:pattern} and S4, there are significant disparities in growth rate of spreading radius in different cities and different time periods. To find intrinsic mechanisms for these disparities, we first divide the spreading period  $T$ ($T$ in Beijing for instance lasted from June 11 to July 10 with 30 days) of each city into two periods $L_1$ and $L_2$ ($L_1$ in Beijing for instance lasted from June 11 to June 25 with 15 days). Then, two spreading radii ($R_1$ and $R_2$) can be calculated based on activity centroids of confirmed cases reported in the spreading periods $L_1$ and $L_2$, respectively. We define the growth rate of spreading radius as $2(R_2-R_1)/|T|$, where $|T|$ denotes number of days in spreading period $T$. Further,  we randomly select 20,000 smartphone users in each city and leverage their trajectory data during the outbreak of COVID-19 to compute their mean travel distance. Since these smartphone users are randomly selected, we use this mean  distance (over 20,000 users) to approximate that of all citizens in  each city. Clearly, a large value of mean travel distance reflects a strong willingness of people for long distance travelling. Considering that different control measures  imposed in Wuhan and Urumqi since the outbreak of COVID-19  affected corresponding mobility pattern and spreading of pandemic significantly (Fig. S7), we divide the spreading period of these two cities into two sub-periods: before and after the implementation of travel restriction, and then calculate mean travel distance and growth rate of spreading radius in these two sub-periods, respectively. The correlation analysis results for all 9 cites are illustrated in Fig. \ref{fig:temporal:pattern}G. Interestingly, we observe a clear positive correlation between mean travel distance and growth rate of spreading radius, indicating that mobility pattern accelerates the urban spreading of COVID-19. 

\subsection*{The Spatial Spreading Pattern}
\begin{figure*}
\centering
\includegraphics[width=17.5cm]{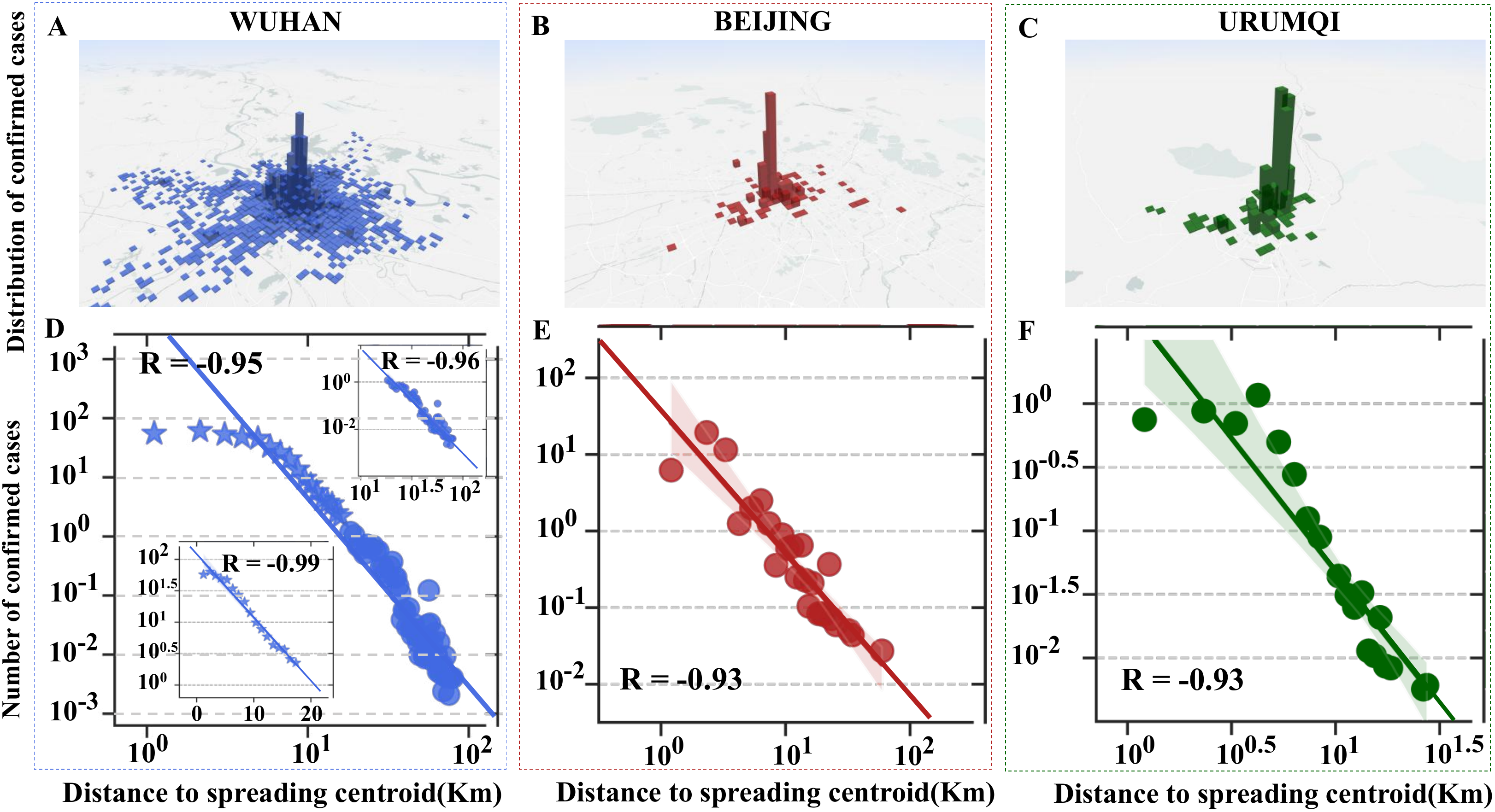}
\caption{The spatial spreading pattern of COVID-19 in Wuhan, Beijing and Urumqi. 
(\textit{A} to \textit{C}) A visualization of the number of confirmed cases in discretized grids in Wuhan, Beijing and Urumqi, respectively. 
(\textit{D} to \textit{F}) The spatial distributions (dots) as a function of distance from overall spreading centroid and the fitted regression lines for these distributions.} 
\label{fig:spatial:pattern-1}
\end{figure*}

To characterize and visualize spatial spreading pattern, we divide the geographical area into grids of $1Km\times 1 Km$\cite{schneckenreither2008modelling}. The overall spreading centroid is set as the original point of grids. Confirmed cases of COVID-19 are then projected into grids according to their activity centroids. As illustrated in Figs. \ref{fig:spatial:pattern-1}A-\ref{fig:spatial:pattern-1}C, three-dimensional histograms are used to describe the spatial distributions of confirmed cases in Wuhan, Beijing and Urumqi, respectively, where the height of each bar represents the case count in each grid.
To analyze the spatial distribution function $F(d)$, we first apply the logarithm to the number of confirmed cases and the distance from overall spreading centroid in all cities. The spatial distributions $F(d)$ of Beijing, Urumqi, Xiaogan, Suizhou and Huanggang  exhibit a prominent linear pattern (Figs. \ref{fig:spatial:pattern-1}E-\ref{fig:spatial:pattern-1}F, Figs. S5 and S6), indicating that the spatial distributions in these cities follow power-law model. Therefore, we adopt $F(d)=d^{\alpha}$ for linear regression. Specifically, we have  $\alpha=-1.80$ for Beijing (the Pearson correlation: -0.93) and $\alpha=-2.15$ for Urumqi (the Pearson correlation: -0.93), respectively. 

As shown in Fig. \ref{fig:spatial:pattern-1}D, the spatial distribution of confirmed cases in Wuhan is power-law-like since it deviates slightly from  power-law model when $d$ is small. Due to the influence of human mobility pattern in Wuhan during the lockdown period (Fig. S7A), initial cases have a higher probability to infect susceptible individuals around spreading centroid. As a result, the risk of infection around spreading centroid is much higher than that at distant locations (a significant 92\% of cases are close to the spreading centroid). To have a more accurate quantification of the spatial distribution of Wuhan, we divide the area into two parts by their distance to the spreading centroid, and adopt two different models to fit the data. Specifically, spatial distribution of confirmed cases around spreading centroid ($d\leq18$) is fitted by an exponential model $F(d)=\alpha^d$, and that when $d>18$ is fitted by a power-law model. As illustrated in Fig. \ref{fig:spatial:pattern-1}D,  $F(d),d\leq18$ is well fitted by an exponential model with Pearson correlation of $-0.99$ and  $F(d),d>18$ is well fitted by a power-law model with Pearson correlation of $-0.96$. This indicates that spatial distribution of confirmed cases in Wuhan indeed has different characteristics, depending on the distance to the spreading centroid. We also observe the similar phenomenon in Xiangyang (Fig. S5F). Interestingly, we find that this is mainly determined by human mobility pattern (to be elaborated in the next Section). 

We also notice that for cities (such as Guangzhou and Wenzhou) where imported cases are widely scattered, the spatial spreading pattern is less prominent. It is clear that there are multiple clusters of confirmed cases in these two cities (Fig. S6), which impacts the power-law-like spatial spreading. Therefore, the observed spreading pattern does not apply to the case with multiple infection sources.

\subsection*{The Underlying Mechanism}
The classic susceptible-infected-recovered model (SIR) and its variants have been widely adopted to understand the transmission characteristics of infectious diseases. The Kendall model \cite{kendall1957discussion,ruan2007spatial,kendall1965mathematics} introduces the spatial dimension to the SIR model and can be used to explain the spatial-temporal evolution of infectious diseases. Its differential equations can be expressed as equations (1)-(3) of supplementary materials. Note that confirmed cases in China get isolated for medical treatment once they are confirmed and would not cause further infection. Under such a condition, the confirmed cases can be regarded as recovered individuals in Kendall model. Then, the differential equation for proportion of recovered individuals can be written as 

\begin{equation}
\begin{split}
\frac{\partial R}{\partial t} &= -\lambda R(x, t)+ \lambda I_{0}(x)
\\ &
\lambda\left[1-\exp \left(-\frac{1}{\lambda} \int_{-\infty}^{\infty} R(y, t) K(x-y) \mathrm{d} y\right)\right].  
\end{split}
\label{eq:recovered:individual}
\end{equation}

\noindent where  $R(x, t)$ denotes the proportion of recovered individuals at location $x$ and time $t$, satisfying $R(x, 0)=0$. Note that $\lambda$ can be obtained by inverse of the basic regeneration number $R_0$ in the model, that is,  $\lambda = 1 / R_0 = \gamma / \beta \xi $, where $\gamma$, $\beta$, $\xi $ represents recovery rate, infection rate and the number of initial susceptible individuals, respectively. Besides, the kernel function $K(x-y) >0$, satisfying $\int_{-\infty}^{\infty} K(y) \mathrm{d} y=1$, quantifies the probability that an infected individual at location $y$ visits $x$. Here, we use power-law distribution to describe the city-level movement behaviors\cite{gonzalez2008understanding}, which can be written as $K(\Delta r)=\Delta r^{\eta}$. Hereby, $K(\Delta r)$  represents the probability for the step size $\Delta r$ and $\eta$, the power-law exponential, denotes  the travel willingness, which has a strong correlation with the mean travelling distance. 

To fit the model for recovered individuals, we first calculate the parameter $\eta$ in the power-law distribution by utilizing the mobility data of anonymous smartphone users, with which to capture the inherent human movement behaviors for each city. Moreover, the diagnosed date for each confirmed case and corresponding activity centroid are also calculated as input of the model. Through the fitting the model (i.e., equation (\ref{eq:recovered:individual})) based on these precalculated parameters and Least Squares algorithm, we can finally obtain a set of optimal parameters ($\lambda$ and  $I_0(x)$). Note that we assume initial confirmed cases originates from the grid $(0,0)$. The parameter $I_0(x)$ fitted in the model could therefore be written as $I_0(0,0)$. A detailed discussion about Kendall model and parameter fitting process are provided in supplementary materials. 

Figs. \ref{fig:model:simulation}A-\ref{fig:model:simulation}C illustrate the evolution of recovered individuals $R(x,t)$ under different timestamps for Wuhan, Beijing and Urumqi, in which Root-Mean-Square-Error (RMSE) is also added to quantify the performance of model fitting.  The values of RMSE for Wuhan, Beijing and Urumqi during whole spreading period are 0.55, 0.18 and 0.57 respectively, indicating that the evolution of recovered individuals $R(x,t)$ during spreading period can be well captured by the proposed Kendall model. More results on the performance of model fitting can be found in Fig. S8. 

We proceed to study the impact of parameter $\eta$ in the model on spatial dispersion of confirmed cases, the number of daily new confirmed cases, and growth rate of spreading radius during the whole spreading period. To characterize the spatial dispersion of confirmed cases, we introduce the concept of Simpson Divergence: $div = 1/\Sigma_i p_{i}^2$, where $p_{i}$ represents the proportion of confirmed cases distributed in grids whose distance to overall spreading centroid is within $[i,i+1]$. Therefore, a small value of $div$ reflects a  high clustering  of the confirmed cases, i.e., a large proportion of confirmed cases are distributed in a small number of grids. Fig. \ref{fig:model:simulation}D illustrates the impact of $\eta$ on spatial dispersion of confirmed cases.
We consider two scenarios under different basic regeneration number $R_0$: $R_0< 1$ and $R_0> 1$.
We see that, with the decrease of $\eta$ Simpson Divergence decreases to 1, at which all confirmed cases are distributed in one grid. The impact of $\eta$ on growth rate of spreading radius is illustrated in Fig. \ref{fig:model:simulation}G. Clearly, the growth rate of spreading radius decreases with the decrease of $\eta$, which is consistent with results in Fig. \ref{fig:model:simulation}D. Figs. \ref{fig:model:simulation}E and \ref{fig:model:simulation}H show the impact of $\eta$ on the number of daily reported confirmed cases. Interestingly, a large $\eta$, which means long mean travelling distance in the human mobility model, results in quick spreading of COVID-19, which makes the peak of daily reported cases arrives early. This also shows that travel restriction policies will  delay the peak arrival. Finally, we study the impact of $\eta$ on the growth rate of spreading radius. As we can see in Figs. \ref{fig:model:simulation}F and \ref{fig:model:simulation}I, spreading radii under different $\eta$  increase with time and converge to a fix value when $R_0>1$. However, when $R_0<1$, the spreading radius increases only when $\eta$ is relatively large. This indicates that when $R_0<1$ and the mean travelling distance is also low, the pandemic will not spread spatially. Therefore, $\eta$ in the mobility model drives the temporal-spatial spreading process. In practice, we can optimize $\eta$ by implementing a specific travel restriction policy to have a desired control and prevention performance.

\subsection*{Discussion}
Most previous studies investigated wide-range (inter-country or inter-state) spreading patterns of infectious diseases\cite{lancet:sun:epidemiological:2020,jia:populationflow:2020,pnas:transportationnetwork:prediction:2006,brockmann2013hidden}, revealing that the number of infected cases at the destination has a strong correlation with total population and the weighted distance from the source to the destination, which is determined by the corresponding population flow. Combining the population flow data and epidemic simulation model, these works accurately characterize large-scale spatial-temporal spreading of epidemics \cite{wang2018characterizing} and predict future spreading trends \cite{hufnagel2004forecast}. Since the population flow is driven by human mobility, existing works have examined the intrinsic mechanism of how human mobility impacts the spreading of diseases\cite{pnas:multiscalemobility:prediction:2009,wesolowski2015impact} and provided theoretical insights about how to mitigate transmission of epidemics through travel restriction\cite{hollingsworth2006will,ferguson2006strategies}. In particular, since the outbreak of COVID-19, many studies have been done to find a more effective and practical way of slowing down the spreading of COVID-19 \cite{science:Kraemereabb4218,tian2020investigation,aleta2020modelling,science:2020:Chinazzieaba9757,lopez2020end}. 
Since the spatial-temporal spreading pattern of infectious diseases is highly dependent on spatial scale\cite{pnas:multiscalemobility:prediction:2009}, previous works focused on the spreading patterns  between countries or states could not be applied to urban scenario directly. 

In this article, we use the data of confirmed cases' activity centroids to study the spatial-temporal spreading pattern of COVID-19 in China. Our results are complementary to previous research, and will provide a fresh perspective to understand the transmission dynamics of COVID-19. Our findings can be used to find the most possible infection center (spreading centroid), evaluate the growth rate and the infection risk of different communities in a new outbreak of COVID-19. Such information is very helpful for practical control and prevention. One shortcoming of the study is that we do not have the information of all confirmed cases in the studied cities. Instead, we only obtain those who use location based services and authorize their activity location data for research purposes. This could bring potential bias to our analysis. To have a valid conclusion, we study the spreading pattern of COVID-19 in 9 cities in China, and the results turn out to be consistent. 
The selection of confirmed cases in our analysis can be viewed as a result of random sampling from all the confirmed cases. Therefore, the results have a good approximation.   

\matmethods{Please see the SI Appendix for detailed information about: 1) Data Description; 2) Evaluation for activity centroid; 3) Sensitivity analysis on temporal spreading pattern of COVID-19; 4) Results of temporal spreading pattern in other 6 cities; 5) Results of spatial spreading pattern in other 6 cities; 6) The mechanism for Spatial-Temporal Spreading Pattern.

\subsection*{Data Availability}
All right of the data utilized in this paper is reserved by Westlake Institute for Data Intelligence.
}



\showmatmethods{} 

\acknow{We thank Jinming Xu for help with the data analyses
of Kendall model and Junfeng Wu for assistance with differential equations calculation. This project was approved by the Ethics Committee of school of Medicine, Zhejiang University}

\showacknow{} 



\clearpage

\clearpage

\clearpage

\clearpage

\clearpage
\begin{figure*}
\centering
\includegraphics[width=17.5cm]{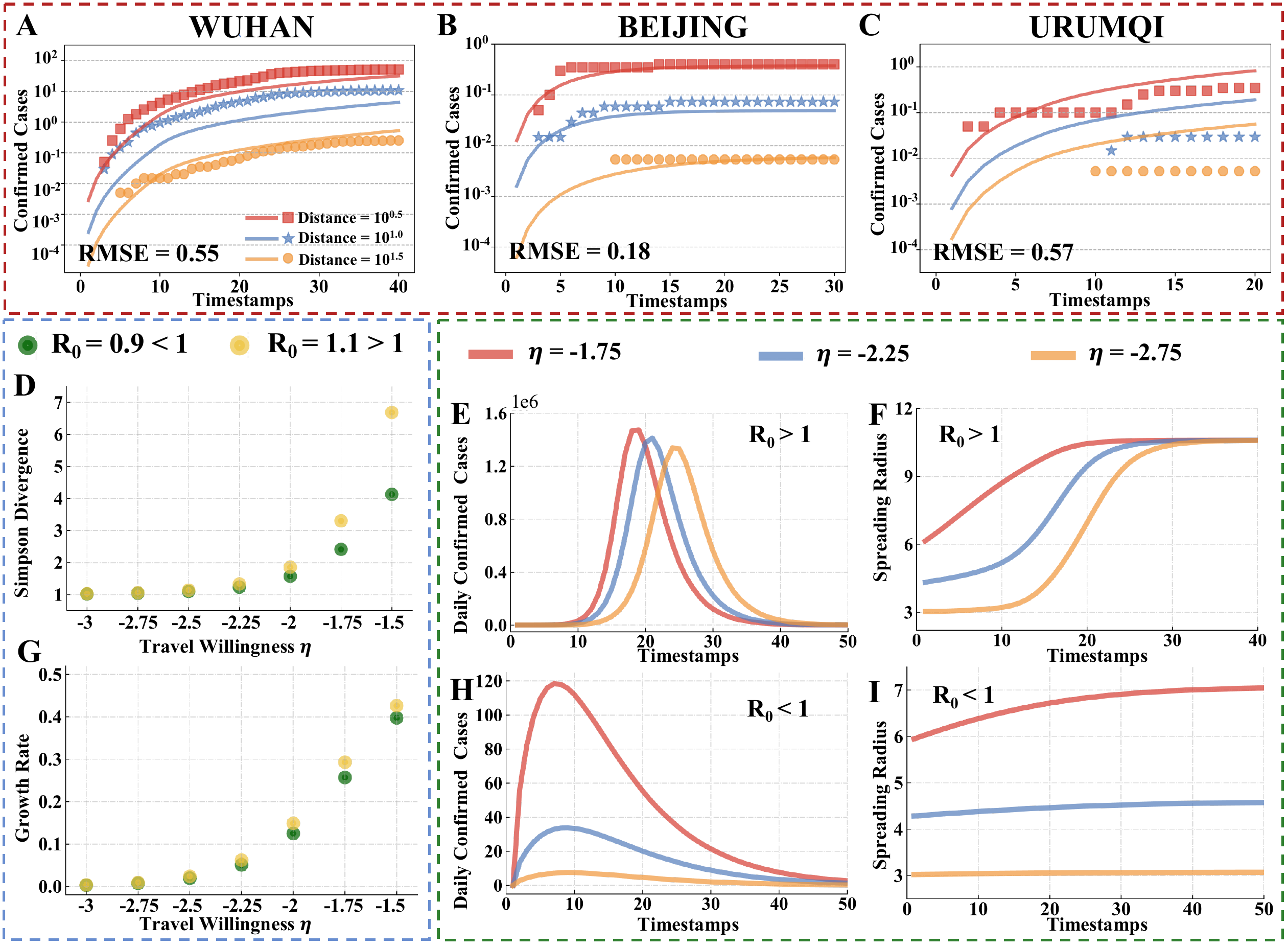}
\caption{The spatial-temporal model for COVID-19 spreading pattern. 
(\textit{A} to \textit{C}) The fitting performance with respective to spatial dimension. 
(\textit{D}) Relation between travel willingness $\eta$ and Simpson divergence.
(\textit{E} and \textit{H}) Relation between travel willingness $\eta$ and daily reported confirmed cases. 
(\textit{G}) Relation between travel willingness $\eta$ and the growth rate of spreading radius.
(\textit{F} and \textit{I}) Relation between travel willingness $\eta$ and spreading radius.}
\label{fig:model:simulation}
\end{figure*} 

\end{document}